\documentclass[letterpaper, 10 pt, conference]{ieeeconf}  

\IEEEoverridecommandlockouts                              

\usepackage{graphicx} 
\usepackage{amsmath} 
\usepackage{amssymb}  
\usepackage{cite}
\usepackage{gensymb}
\usepackage{comment}

\title{\LARGE \bf
Experimental Comparison of Hardware-Amenable Spike Detection Algorithms for iBMIs
}

\author{Shoeb Shaikh$^{1}$, Rosa So$^{2}$, Camilo Libedinsky$^{3}$ and Arindam Basu$^{1}$
\thanks{**This work was supported through grant RG87/16 by MOE, Singapore}
\thanks{$^{1}$Authors are with Nanyang Technological University, Singapore
        {\tt\small arindam.basu@ntu.edu.sg}}%
\thanks{$^{2}$Author is with the Institute for Infocomm Research, Singapore
        {\tt\small rosa-so@i2r.a-star.edu.sg}}
\thanks{$^{3}$Author is with SiNAPSE, National University of Singapore, Singapore
        {\tt\small camilo@nus.edu.sg}}%
}

\begin{document}
\maketitle
\thispagestyle{empty}
\pagestyle{empty}

\begin{abstract}

This paper presents an experiment based comparison of absolute threshold (AT) and non-linear energy operator (NEO) spike detection algorithms in Intra-cortical Brain Machine Interfaces (iBMIs). Results show an average increase in decoding performance of  $\approx$ 5\% in monkey A across 28 sessions recorded over 6 days and $\approx$ 2\% in monkey B across 35 sessions recorded over 8 days when using NEO over AT. To the best of our knowledge, this is the first ever reported comparison of spike detection algorithms in an iBMI experimental framework involving two monkeys. Based on the improvements observed in an experimental setting backed by previously reported improvements in simulation studies, we advocate switching from state of the art spike detection technique - AT to NEO.

\end{abstract}

\section{INTRODUCTION}

Intra-cortical Brain Machine Interfaces (iBMIs) have enabled patients suffering from debilitating spinal cord afflictions to regain a sense of control through demonstrations such as \cite{Collinger2013, Ajiboye2017} where the subjects are successfully able to feed themselves and \cite{Pandarinath2017} where record communication speeds have been achieved among other things \cite{Libedinsky2016b, Rajangam2016}. Spikes serve as an input to these systems, where instantaneous firing rates are computed on every recorded electrode and subsequently mapped to a behavioral co-variate to drive an effector or stimulate a paralysed limb.

The process of spike detection entails identifying a spike amidst the recorded background noise. Absolute Threshold (AT) method is the state of the art technique currently employed for this purpose in both non-human primate (NHP) \cite{Gilja2012, Libedinsky2016b, Christie2015} and human subject \cite{Pandarinath2017, Ajiboye2017} based studies. However, this technique is sensitive to placement of an optimal threshold. If one sets it too high, spikes are missed and if one sets it too low, false detections occur \cite{Rey2015}. Secondly, hardware development in iBMIs is progressing towards including spike detection in the implant itself to counter the problem of transmitting broadband raw data rate \cite{Chattopadhyay2017, Chen2016, Boi2016, Harrison2007}. For such a design decision, it is not feasible to implement AT on chip as it requires large memory and concomitantly large area for computation of a median-based threshold \cite{Wing-kinTam2015}.
\begin{figure}[t]
\centering
\includegraphics[scale=0.5]{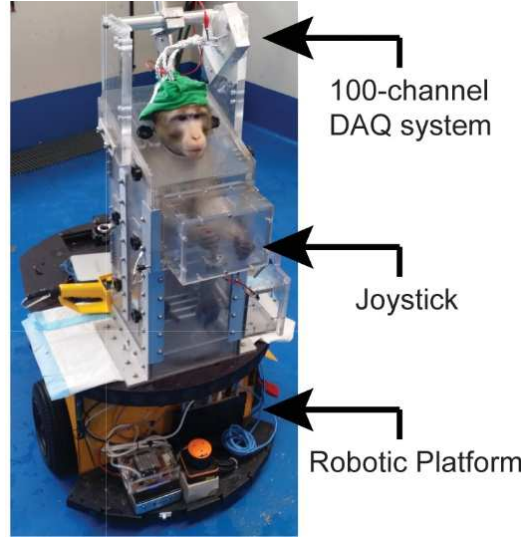}
\caption{Experimental setup for the iBMI experiment adapted from \cite{Libedinsky2016b}}
      \label{fig:expt}
\end{figure}

An alternate technique by the name of non-linear energy operator (NEO) \cite{Teager90} has been shown to perform better in spike detection than AT in studies based on semi-synthetic data such as \cite{Mukhopadhyay1998, Gibson2010}. Apart from being better adept at spike detection due to reasons which will be briefly touched upon in $Section$ $II$, NEO enjoys two specific advantages over AT. Firstly, it has been shown to be generally less sensitive to the placement of threshold as compared to AT \cite{Gibson2010}. Secondly, it is computationally friendly without any demanding memory requirements and extreme low power implementations such as \cite{Detector2016, Yang2017}  consuming $40$ nW/channel, $50$ nW/channel respectively. Other spike detection techniques such as stationary wavelet transform product \cite{kim2003}, exponential component - power component (EC-PC) \cite{ZhiYang2012} have been reported in literature, but these were not considered in this analysis owing to their relatively greater computational complexity \cite{Gibson2010, Zou2014} than both AT and NEO.

Spike detection methods reported so far in literature are based on studies performed on synthetic and semi-synthetic data with realistic assumptions \cite{Rey2015}. The logical next step we believe is to apply these findings in a behavioral task-based experimental framework to check for their validity. An analysis along similar lines has been reported in \cite{Wing-kinTam2015}. However, this study lacks on two fronts. Firstly, it is restricted to just one NHP instead of the standard protocol of two NHPs. Secondly, the proposed method's hardware implementation consumes up to three orders of magnitude more power than the ones reported for NEO \cite{Zou2014, Detector2016, Yang2017} reflecting it's computational complexity.

Thus, the object of this study is to apply the simulation based finding of NEO's preponderance over state of the art - AT in a simple experimental setting for two NHPs with iBMI performance as a yardstick. The line of reasoning is similar to \cite{Christie2015}, where spike sorting and thresholding were compared in context of iBMI performance.

\section{Spike Detection Algorithms}

In this section, we briefly present AT and NEO spike detection methods.

\subsection{Absolute threshold}

This method involves applying a threshold to the digitally filtered raw data $x[n]$ at time $n$ to identify a spike. Threshold is computed over $N$ samples of filtered raw data as,

\begin{IEEEeqnarray}{CCl}
$$\sigma_N = \frac {median|x[n]|}{0.6745} \label{eq:1}
\\
Thr_{AT} = k \times \sigma$$
\end{IEEEeqnarray}where $\sigma_N$ is the standard deviation estimate over $N$ samples of filtered raw data, $k$ is an integer typically between $2$ and $5$. Negative threshold crossings of the filtered raw signal are considered as spikes. 

\subsection{Non-linear Energy Operator}

This method involves application of a pre-emphasis step before spike detection \cite{Mukhopadhyay1998}. The idea behind this approach is to accentuate high energy, high frequency content amidst background noise thereby increasing spike to noise separation. NEO threshold calculation involves the following steps,

\begin{IEEEeqnarray}{CCl}
\psi [n] = x^2[n] - x[n-1] \times x[n+1]  
\\
Thr_{NEO} = l \times \frac {\sum_{n = 1}^{N} \psi [n]}{N} 
\end{IEEEeqnarray}where $\psi [n]$ is the NEO processed, pre-emphasised signal at time $n$. $Thr_{NEO}$ is then computed as a multiple - $l$ times the average of $\psi [n]$ over $N$ samples of filtered raw data. The value of $l$ is optimized as per the experimental conditions \cite{Gibson2010, Zou2014} and is not as sensitive as AT due to better spike to noise separation. Other methods of automatic threshold setting that are less sensitive to amplitude of the NEO output can also be used \cite{Detector2016}. Also, NEO can be applied directly to the analog signals after the amplifier and before the ADC to save the ADC power dissipation \cite{Detector2016}. 

\begin{figure}[h]
\centering
\includegraphics[scale=0.26]{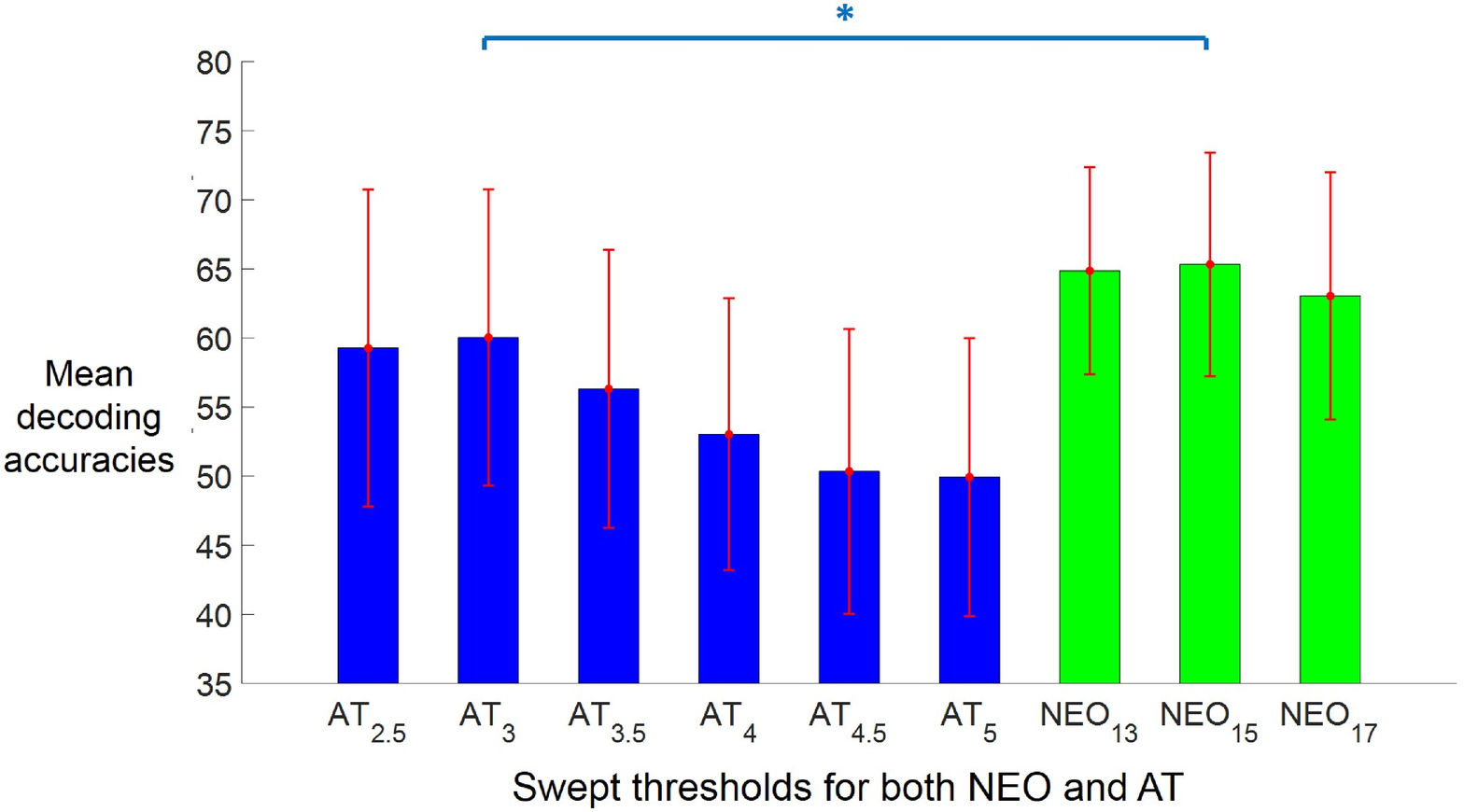}\\
(a)
\includegraphics[scale=0.26]{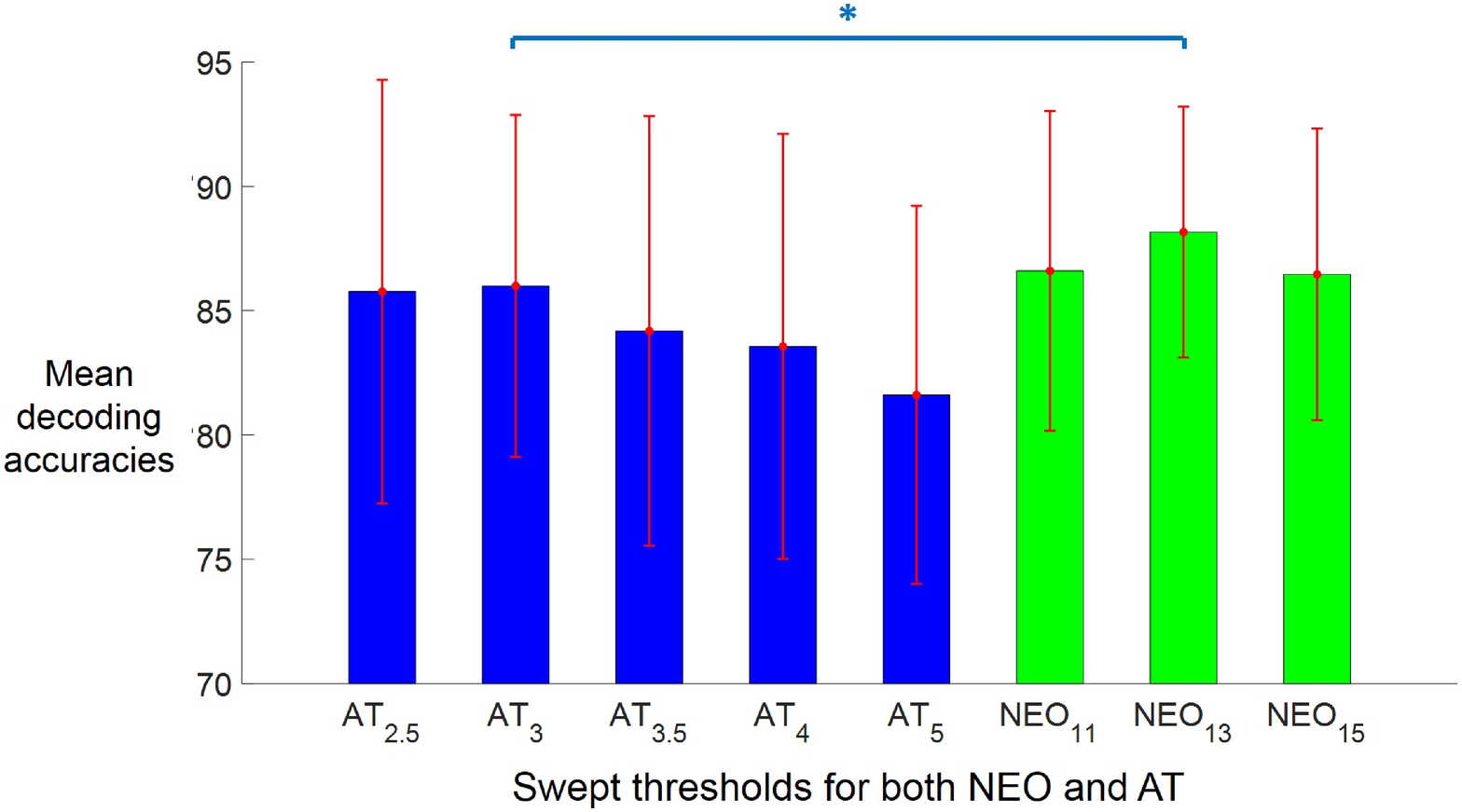}\\
(b)
\caption{Decoder results for (a) monkey A and (b) monkey B. Statistical significance test is conducted only between best decoding performance of AT and NEO, significant results are denoted by (*), p$<$0.05 (paired t-test). Higher decoding accuracies are obtained using NEO in both cases. Threshold multiples $k$ and $l$ for AT and NEO respectively are marked in the figure for each bar.}
      \label{fig:monkAres}
\end{figure}

\section{METHODS}

\begin{figure*}[t]
\centering
\includegraphics[scale=0.5]{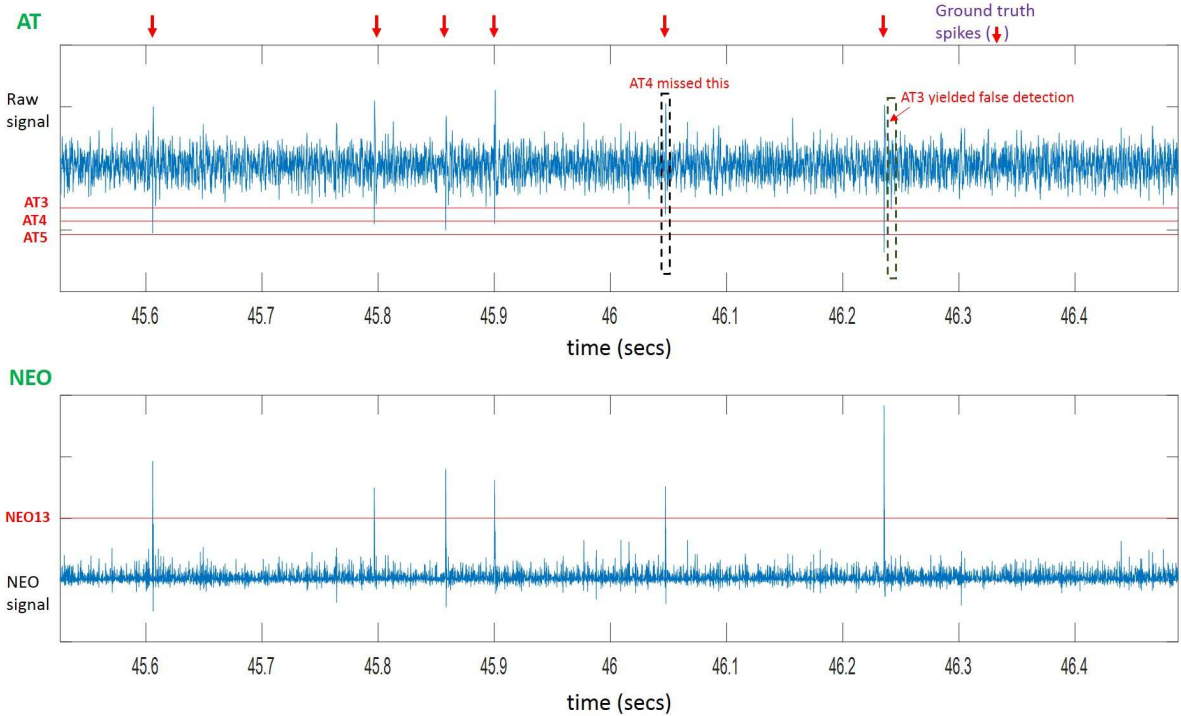}
\caption{NEO vs AT visualization for a segment of neural recording showing improved detection using NEO.}
      \label{fig:visualize}
\end{figure*}
\subsection{Neural Data Acquisition} 
Multiple microelectrode arrays were implanted in the hand/arm area of primary motor cortex. There were a total of $96$ and $64$ implanted electrodes sensing extracellular neural signals in monkeys A and B respectively. These signals were then passed through front-end circuitry sampling processed raw data at $13$ $k$Hz \cite{Zou2013}. The digitized raw data so obtained was subjected to digital bandpass filtering between $300$ and $3000$ Hz implemented on a PC to yield $x[n]$ at time $n$.  

\subsection{Experiment}

The experiment protocol is described in detail in \cite{Libedinsky2016b}. Briefly, a monkey was trained to manipulate the position of a robotic platform through a hand-controlled three-directional joystick. Four tasks - moving forward by $2$ m, turning $90\degree$ right, turning $90\degree$ left, staying still for $5$ seconds constituted the experiment. Trial was considered successful if the monkey reached the target within $15$ seconds for tasks involving movement and stayed still for $5$ seconds during the stop task. A successful trial was rewarded with a fruit reward by the trainer.

\subsection{Decoding Algorithm}
We have used linear discriminant analysis (LDA) as the decoder of choice for mainly two reasons. Firstly, it is popularly used in BMIs operated in a discrete control fashion \cite{Libedinsky2016b, Lotte2007, Lotte2018}. Secondly, LDA is a relatively simpler algorithm compared to more complex algorithms such as support vector machine and deep neural nets which have in some cases shown to outperform simple linear algorithms \cite{Lotte2018}. This ensures that the gains in decoding accuracy are primarily attributed to the spike detection method and not the superior generalization capabilities of the classification algorithm \cite{Wing-kinTam2015}.

\section{RESULTS}
The LDA based decoder was trained de novo on the first session's data on each day. Subsequent $3$-$4$ sessions were used as test sessions for offline analysis. A total of $28$ sessions across $6$ days and $35$ sessions across $8$ days were used in the analysis for monkeys A and B respectively. Thresholds for both AT and NEO methods were set using the initial 30 seconds of raw data from the first session. For a fairer comparison multipliers $k$ and $l$ described in equations (2), (4) were swept for both AT and NEO respectively. 

The averaged decoder results across all the test sessions are plotted in figures \ref{fig:monkAres}(a) and (b) respectively. For monkey A, NEO with $l = 15$ ($NEO_{15}$) and AT with $k = 3$ ($AT_{3}$) perform the best respectively, and $NEO_{15}$ outperforms $AT_{3}$ by $5.29$ \%. Along similar lines for monkey B, NEO with $l = 13$ ($NEO_{13}$) and AT with $k = 3$ ($AT_{3}$) perform the best respectively, and $NEO_{15}$ outperforms $AT_{3}$ by $2.16$ \%. Another point to note is that in both monkey cases, standard deviations of decoding accuracies across sessions for NEO are less than AT. For monkey A, the standard deviation of decoding accuracies across sessions for $NEO_{13}$ is $8.09$ \%, whereas for $AT_{3}$ it is $10.72$ \%. Similarly for monkey B, standard deviation of decoding accuracies across sessions for $NEO_{13}$ is $5.04$ \%, whereas for $AT_{3}$ it is $6.88$ \%. Thus, NEO performs better than AT with greater average and lower standard deviation of decoding accuracies across sessions for both monkeys.


\section{DISCUSSION}
Initial iBMI studies resorted to spike sorting to map cortical neural firing rates to behavioral variables \cite{Serruya2002, Taylor2002}. However, studies such as \cite{Christie2015, Fraser2009} have shown that perhaps it is unnecessary, with little or no loss in decoding accuracies obtained by forsaking sorting for simple threshold crossings. Furthermore, there have also been studies to provide a quantitative understanding for using threshold crossing in lieu of sorted spikes \cite{Trautmann2017b}. Thus, there is pressing evidence to simply use threshold crossings to monitor the population level response as a whole from intra-cortical microelectrode recordings. We argue that, NEO by virtue of being better adept at spike detection yields a more informative population response. 

In figure \ref{fig:visualize}, we have visualized filtered raw data and NEO processed signal along with different thresholds for NEO, AT for a single channel as a way to demonstrate the difference in spike detection between these two approaches. Ground truth spikes are marked in the form of downward pointing arrows. One can see that for AT with $k = 4$, $k = 5$, the threshold levels are relatively higher and some spikes are missed being detected. One might be tempted to go for a further lower threshold with $k = 3$ to rein in the missed detections. However, with $k = 3$, some false detections can be seen to occur. Simple eyeballing in the lower half of the plot for NEO with $l = 13$ reveals that spikes are comfortably detected with relatively lesser concern for false detection. One can easily observe the increased spike to noise ratio advantage offered by NEO over AT in figure. \ref{fig:visualize}.

\begin{figure}[h]
\centering
\includegraphics[scale=0.4]{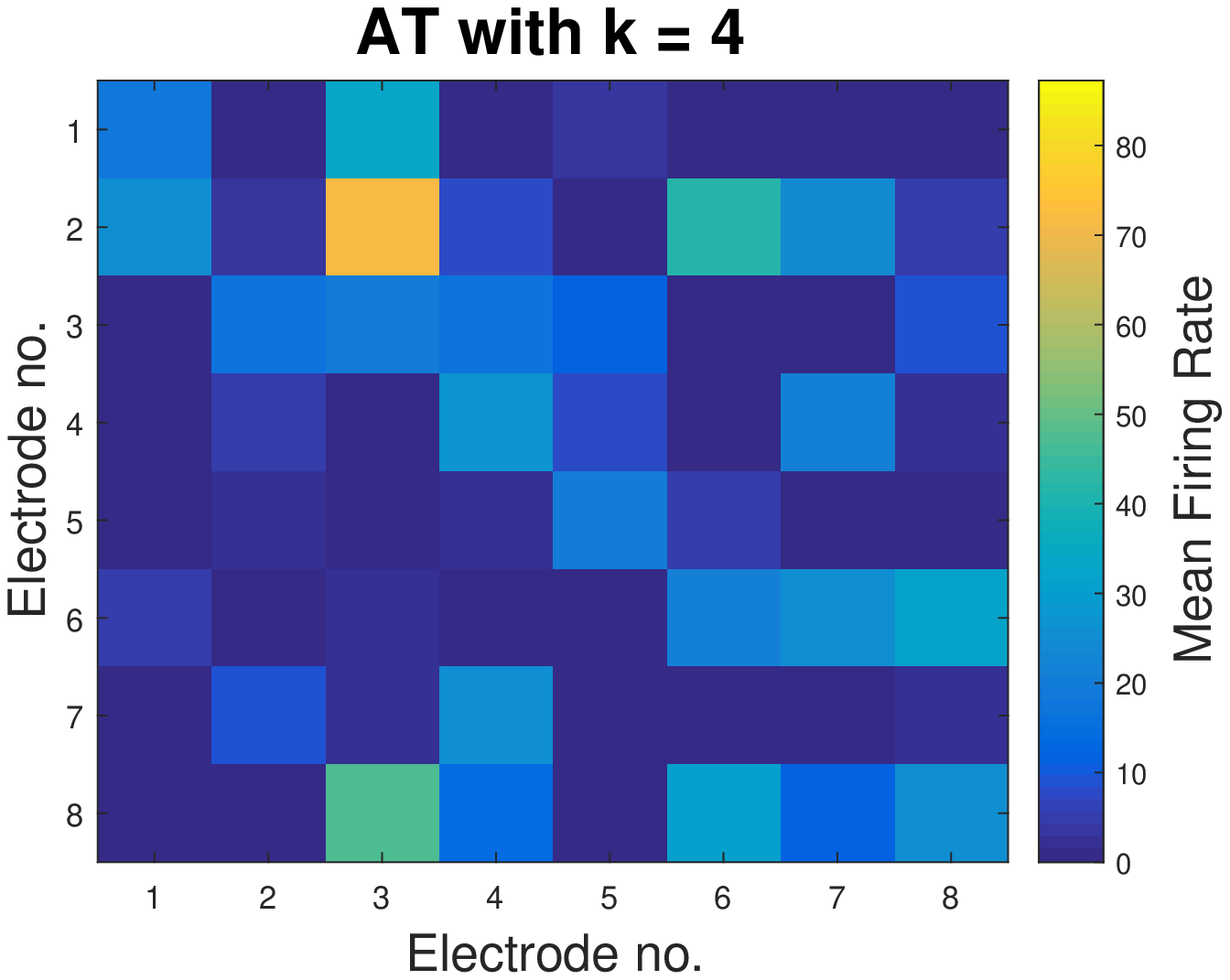}\\
(a)\\
\includegraphics[scale=0.4]{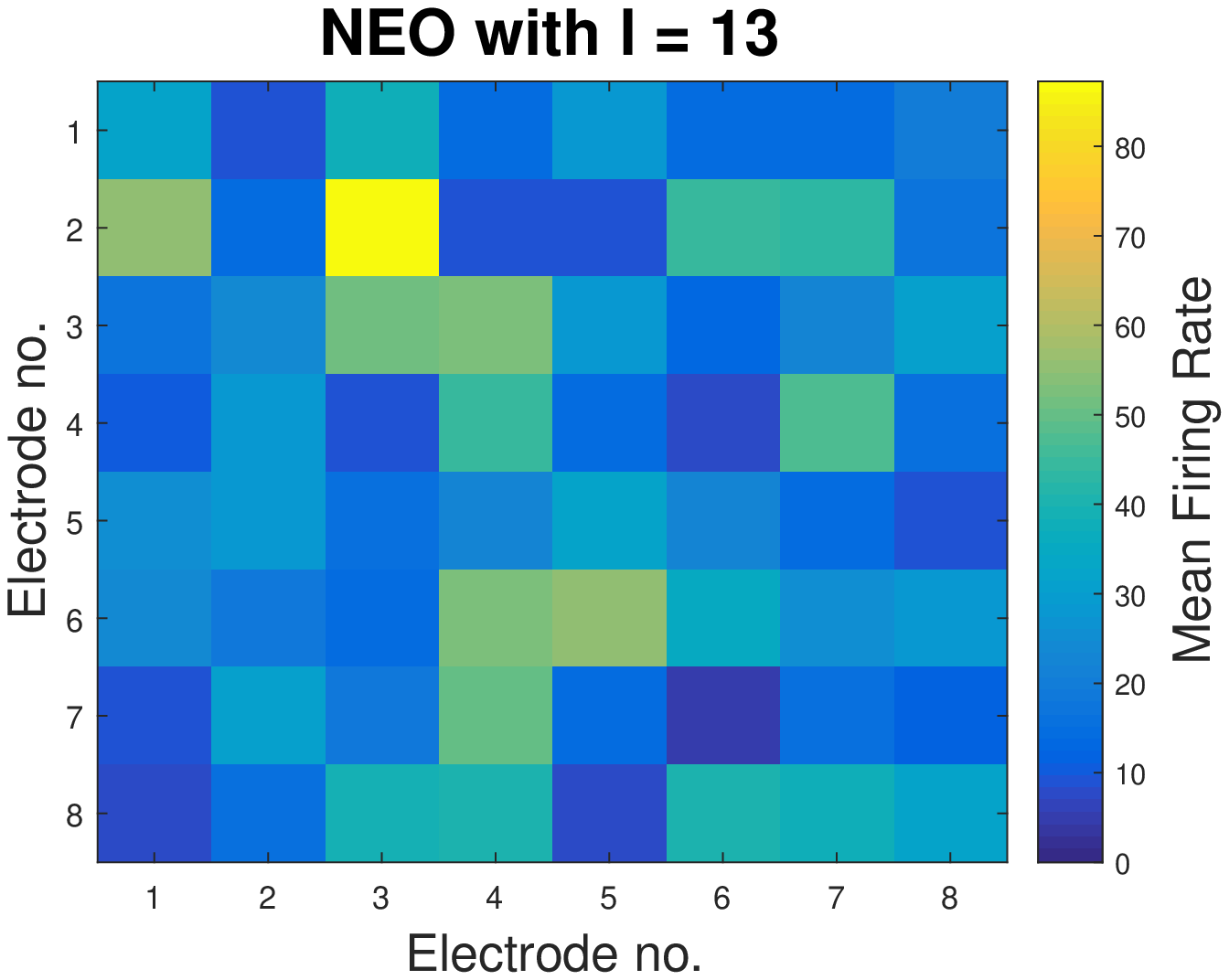}\\
(b)
\caption{Mean firing rates on electrodes (plotted in an $8 \times 8$ array fashion) with (a) AT ($k = 4$) and (b) NEO ($l = 13$) for Monkey B when it is performing the forward trials in a given session}
      \label{fig:AT4}
\end{figure}

The color-coded firing rate plots in figures \ref{fig:AT4}(a) and (b) also show evidence of increased firing rates appearing at the electrodes for NEO with $l = 13$ as compared to AT with $k = 4$. This goes to show that NEO helps in detecting spikes from neurons which were perhaps missed by AT and helps in gleaning increased population-level information.

\section{CONCLUSION}
To conclude, we have shown NEO to be better than AT in context of iBMI performance for two monkeys with $\approx 2\%$ and $\approx 5\%$ improvement in decoding accuracy respectively. The underlying hypothesis is that NEO helps in better spike detection which in turn leads to enhanced performance. Thus, based on previous simulation studies \cite{Mukhopadhyay1998, Gibson2010}, current results and hardware amenability \cite{Detector2016} we advocate switching from AT to NEO as the de facto spike detection method. Future work includes closed-loop trials and applying NEO in conjunction with more sophisticated decoders such as neural networks and Kalman filters.

\section*{ACKNOWLEDGMENT}
The authors would like to thank Clement Lim for helping in animal training and data collection.

\bibliographystyle{IEEEtran}

\end{document}